# Achieving Near-Capacity at Low SNR on a Multiple-Antenna Multiple-User Channel

Chau Yuen                Bertrand M. Hochwald

cyuen@i2r.a-star.edu.sg       hochwald@beceem.com

**Abstract**

We analyze the sensitivity of the capacity of a multi-antenna multi-user system to the number of users being served. We show analytically that, for a given desired sum-rate, the extra power needed to serve a subset of the users at low SNR (signal-to-noise ratio) can be very small, and is generally much smaller than the extra power needed to serve the same subset at high SNR. The advantages of serving only subsets of the users are many: multi-user algorithms have lower complexity, reduced channel-state information requirements, and, often, better performance. We provide guidelines on how many users to serve to get near-capacity performance with low complexity. For example, we show how in an eight-antenna eight-user system we can serve only four users and still be approximately 2 dB from capacity at very low SNR.

## 1. INTRODUCTION

Given a base-station or access-point with $M$ transmit antennas serving a pool of $L$ autonomous single-antenna users who cannot cooperate with each other, we wish to approach the sum-capacity, especially at low SNR. This is a multi-antenna multi-user communication system, which is also sometimes referred to as a Gaussian vector broadcast channel. The sum-capacity of such a system is investigated in [1-6], assuming that channel state information (CSI) is available at the transmitter and receivers.

Achieving capacity in this multi-antenna multi-user channel ostensibly could in theory require us to serve all $L$ users simultaneously with the $M$ transmit antennas. Generally, most systems have $L \geq M$. For such a system, the sum-rate of serving $K$ users (out of the $L$) grows linearly with $K$ for $K \leq M$ and sub-linearly for $K > M$ [7]. Since system complexity can grow significantly with the number of users served, an advantageous performance-complexity tradeoff can be attained by serving $M$ users. In fact, we show that we may often serve fewer than $M$ users with little rate penalty, especially at low SNR. We quantify the penalty in this paper.

We focus on low SNR's or low SINR's (signal to interference-plus-noise ratio) because many multi-user systems operate in an interference-dominated environment. Many algorithms have been proposed for MIMO multi-user. However, some multi-user techniques, such as vector-perturbation described in [8], are able to approach the sum-capacity at high SNR but suffer some performance loss at low SNR. It is then important to improve the performance of these algorithms at low SNR. One simple way to reduce complexity and improve per-user performance is simply to serve fewer users. But serving fewer users is a viable strategy only if we do not suffer a large total throughput penalty in the process. We quantify the penalty and show that it is small at low SNR's.

We define and use a quantity called *sensitivity* of the capacity as follows: Given that we are serving a certain number of users at a certain SNR and total throughput, if we now serve fewer users (using the same number of transmitter antennas), how much do we need to increase the SNR to obtain the same total throughput? The reduction in number of users is the equivalent of a complexity reduction (that comes from lower channel-state information requirements and simplified algorithms), while the increase in the SNR is the equivalent of a power penalty. For example, for eight transmit antennas, at $\rho = 1$ dB (which is defined as the ratio of total transmit power to per-user receive noise power), we suffer a penalty of only 1.1 dB in SNR (increase in $\rho$) if we serve only four users chosen randomly, each at one-fourth the sum-rate, versus all eight users, each at one-eighth the rate. The corresponding penalty at $\rho = 10$ dB is more than 2.55 dB, and the corresponding penalty at $\rho = 20$ dB is more than 7.63 dB. We sometimes call this power penalty a "loss" and this loss is clearly much lower at small $\rho$ in this example. Other definitions of capacity sensitivity are also possible, such as in [11] where the suboptimality of TDMA at low power is quantified.

Thus, to obtain high rates at low SNR's we may as well serve fewer users with algorithms that have low complexity and good performance. Continuing the example of the previous paragraph, we halve the number of users being served and suffer a 1.1 dB penalty in the SNR at $\rho = 1$ dB. In return, we obtain the benefit of requiring the channel state information (CSI) of only four users at the transmitter. Furthermore, we may obtain performance gains from any algorithm used to serve the users: for example, when serving four users, the vector-perturbation technique gains approximately 1.5 dB over the same technique serving eight users (in the eight-antenna system of this example). Hence the algorithmic gain of serving only four users exceeds the power penalty and we have a net gain.

The user selection process can be fair. Even if we choose to serve four users rather than eight at each transmission in the above example, we require the total sum-rate to be the same. Therefore, each user gets twice as much data per transmission as when all eight are served. A different set of four users may be chosen at each transmission, and all users can therefore be served in round-robin fashion with no loss in rate.

Rather than work directly with the capacity, which is often intractable, we generally work with a tractable lower bound that gives us insight into the capacity. We formulate the problem in detail in the next section.

## 2. PROBLEM FORMULATION

Consider a multi-antenna multi-user communication system with $M$ transmit antennas and a pool of $L$ users, each with one antenna. The sum-capacity is [1]

$$C_L = \mathrm{E} \max_{\mathbf{D}_L, \mathrm{tr}(\mathbf{D}_L)=1} \log \det\left(\mathbf{I}_M + \rho \mathbf{H}^* \mathbf{D}_L \mathbf{H}\right) \tag{1}$$

where $\mathbf{I}_M$ is an $M \times M$ identity matrix, $\mathbf{H}$ is the $L \times M$ is the channel matrix between every user and the transmitter, and $\mathbf{D}_L$ is an $L \times L$ positive diagonal matrix whose trace is unity. The elements of $\mathbf{H}$ are Rayleigh fading (complex-Gaussian) coefficients with zero mean and unit variance. Throughout this paper, log is base-two, and natural log is denoted ln. We ignore the numerical and algorithmic issues of optimizing $\mathbf{D}_L$ in (1).

The sum-capacity (1) requires full knowledge of the CSI at the transmitter and hence may be difficult to achieve if $L$ is large. We consider the case of serving $M$ users (with $M$ antennas) chosen *randomly* from the pool of $L$. In this case the achievable sum-rate is

$$I_M = \mathrm{E} \max_{\mathbf{D}_M, \mathrm{tr}(\mathbf{D}_M)=1} \log \det\left(\mathbf{I}_M + \rho \mathbf{H}^* \mathbf{D}_M \mathbf{H}\right) \tag{2}$$

where $\mathbf{H}$ is $M \times M$ complex-Gaussian channel matrix. Achieving this rate requires knowledge of the CSI of any $M$ users at the transmitter. Clearly $C_L \geq I_M$.

The focus of this paper is low SNR. Multiple transmit antennas can still be beneficial at low SNR when channel information is available to the transmitter. This can be justified as follows: when $\rho$ is very small, $I_M$ can be lower bounded by

$$I_M = (\log e) \mathrm{E} \max_{\mathbf{D}_M, \mathrm{tr}(\mathbf{D}_M)=1} \mathrm{tr}\left(\rho \mathbf{H}^* \mathbf{D}_M \mathbf{H}\right) + O(\rho^2)$$
$$\approx (\log e) \rho \, \mathrm{E} \max_{\mathbf{D}_M, \mathrm{tr}(\mathbf{D}_M)=1} \mathrm{tr}\left(\mathbf{D}_M \mathbf{H}\mathbf{H}^*\right) \geq (\log e) \rho M \quad (3)$$

Equation (3) suggests that we should choose $\mathbf{D}_M$ to have a one on its diagonal corresponding to the user whose diagonal entry of $\mathbf{H}\mathbf{H}^*$ is largest; it is therefore best to serve only the user with the largest channel gain. Nevertheless, the inequality is obtained by setting $\mathbf{D}_M = \mathbf{I}_M$ and we see that the sum-rate grows at least linearly with the number of antennas $M$ at low SNR.

We next investigate how closely we can achieve $I_M$ while reducing the number of users served $K$ to a value less than $M$. This would allow us to enjoy the benefits of having $M$ antennas without having to serve $M$ users. The achievable rate of serving $K$ users (with $M$ antennas) is denoted $I_K$. We measure our ability to achieve $I_M$ by computing, for a given $\rho$, how much we need to increase $\rho$ to make $I_K = I_M$ for $K < M$. We use differential analysis to make the computation tractable.

## 3. SENSITIVITY ANALYSIS OF SUM-RATE

### 3.1 Bounding the sum-rate

We consider our ability to achieve $I_M$ in equation (2) as a function of the number of users served. To simplify the analysis, suppose in (2) that we set

$$\mathbf{D}_M = \frac{1}{M} \mathbf{I}_M \quad (4)$$

(a diagonal matrix with $1/M$ on the diagonals) and obtain the following lower bound:

$$I_M \geq I_{\mathrm{eq},M} \quad (5)$$

where
$$I_{\mathrm{eq},M} = \mathrm{E} \log \det \left( \mathbf{I}_M + \frac{\rho}{M} \mathbf{H}^* \mathbf{H} \right) . \quad (6)$$

The suffix "eq" denotes the fact that we are setting $\mathbf{D}_M$ to have equal entries. We use $I_{\mathrm{eq},K}$ to denote serving $K$ users where $K$ can be less than $M$.

The diagonal elements of $\mathbf{D}_M$ in (4) are related to the power assigned to different data streams of different users [1]. Rather surprisingly, we show that the lower bound in (6) (which assigns equal power to the data streams) is tight at both high and low $\rho$ for large $M$. When $M = 1$, this lower bound is trivially tight. For large $\rho$ it is shown in [2] that setting $\mathbf{D}_M = (1/M)\mathbf{I}_M$ leads to a tight bound for any $M$. When $\rho$ is small, we show in Appendix A that

$$(1+\zeta)I_{eq,M} \geq I_M \geq I_{eq,M} \tag{7}$$

for large $M$ and any $\zeta > 0$. In fact, we have found that the rate $I_M$ in (2) and the lower bound $I_{eq,M}$ (6) are often nearly equal for all SNR's.

We determine the sensitivity of the achievable rate to the number of users served by decreasing the number of users by a small amount and examining how much we must increase the SNR so as to keep the total rate constant. Instead of working directly with $I_M$ (which does not have a closed-form), we use the lower bound (5) and the general relationship is as follows:

$$\begin{array}{c} I_M \gtreqless I_{eq,M} \\ \vee \quad \vee \\ I_K \gtreqless I_{eq,K} \end{array} \tag{8}$$

where $K$ is the design variable representing the number of users we serve with $M$ antennas. We would like to find the difference between $I_M$ and $I_K$ as a function of $K$ and $\rho$. We examine this difference indirectly as shown in equation (8) by instead examining the differences in the two circled inequalities. As shown in equation (7), we know that $I_{eq,M}$ is a good approximation of $I_M$. Thus, we only need to investigate the difference between $I_{eq,M}$ and $I_{eq,K}$. If the difference between $I_{eq,M}$ and $I_{eq,K}$ is small, then the difference between $I_M$ and $I_K$ is also small.

### 3.2  Differential analysis of sum-rate

When we reduce the number of users from $K$ to $K'$, the ratio $\beta$ increases to $\beta'$, where $\beta' = M/K'$. In order to achieve the same rate before this reduction, $\rho$ must be increased to some $\rho'$. We define two quantities: $\varepsilon$ the *complexity reduction coefficient,* and $\delta$ the *power penalty coefficient*. The quantity $\varepsilon$ is defined as:

$$\varepsilon = \frac{d\beta}{\beta} \quad \text{where} \quad d\beta = \beta' - \beta \tag{9}$$

A large positive $\varepsilon$ implies a large reduction in complexity. We are interested in infinitesimal changes, and an infinitesimal change in $\beta$ is related to an infinitesimal change in $K$ through

$$\varepsilon = \frac{d\beta}{\beta} = -\frac{dK}{K}. \tag{10}$$

Observe that $\varepsilon$ divides the change in $\beta$ by $\beta$. Hence, if $\varepsilon = 1$ the number of users is halved ($\beta' = 2\beta$). This is a notational convenience since it turns out that our final results are insensitive to the absolute number of antennas and users but are strong functions of the ratio $\beta$.

The power penalty coefficient $\delta$ is defined as:

$$\delta = \frac{d\rho}{\rho} \quad \text{where} \quad d\rho = \rho' - \rho \tag{11}$$

A large positive $\delta$ implies a large increase in SNR. The penalty $\delta$ is related to the dB-change in power through

$$\rho' = \rho(1+\delta) \Rightarrow d\rho_{(\text{dB})} = 10\log_{10}(1+\delta) \approx 10(\log_{10} e)\delta \tag{12}$$

We define the *sensitivity* as the ratio

$$\frac{\delta}{\varepsilon} = \frac{d\rho/\rho}{d\beta/\beta} = \frac{d\rho}{d\beta}\left(\frac{\beta}{\rho}\right) \tag{13}$$

where the changes in $\rho$ and $\beta$ are infinitesimal and such that $I_{\text{eq},K}$ is kept constant. A small value for this ratio suggests that the number of users can be changed (while keeping the rate constant) with little penalty in power.

To obtain the sensitivity we solve

$$I_{\text{eq},K}(\rho) = I_{\text{eq},K'}(\rho') = \text{constant} \tag{14}$$

for infinitesimal changes in $\beta$ and $\rho$.

### 3.3 Formula for sensitivity

The quantity $I_{\text{eq},K}$ has the big advantage of an approximate closed-form formula:

$$\begin{aligned}
I_{\text{eq},K} &= \text{E}\log\det\left(\mathbf{I}_M + \frac{\rho}{K}\mathbf{H}^*\mathbf{H}\right) \\
&= K\text{E}_\lambda \log\left(1+\frac{\rho}{K}\lambda\right) \approx K\text{F}(\beta,\rho)
\end{aligned} \tag{15}$$

Where $\lambda$ denotes any eigenvalues of $\mathbf{HH}^*$, and $\text{F}(\beta,\rho)$ is defined in [9] as:

$$\begin{aligned}
\text{F}(\beta,\rho) &= \frac{1}{\pi}\int_{(\sqrt{\beta}-1)^2}^{(\sqrt{\beta}+1)^2} \log(1+\rho\lambda)\sqrt{\frac{\beta}{\lambda}-\frac{1}{4}\left(1+\frac{(\beta-1)}{\lambda}\right)^2}\,d\lambda \\
&= \log\left(1+\rho\left(\sqrt{\beta}+1\right)^2\right) + (\beta+1)\log\left(\frac{1+\sqrt{1-a}}{2}\right) \\
&\quad -(\log e)\sqrt{\beta}\,\frac{1-\sqrt{1-a}}{1+\sqrt{1-a}} + (\beta-1)\log\left(\frac{1+\gamma}{\gamma+\sqrt{1-a}}\right)
\end{aligned} \tag{16}$$

where $a = \dfrac{4\rho\sqrt{\beta}}{1+\rho\left(\sqrt{\beta}+1\right)^2}$, $\gamma = \dfrac{\sqrt{\beta}-1}{\sqrt{\beta}+1}$, and where $\mathbf{H}$ is of dimension $K \times M$ and $\beta = M/K$.

The approximation of $I_{eq,K} \approx K\mathrm{F}(\beta, \rho)$ becomes an equality when we consider a fixed $\beta$ but both $K$ and $M$ are allowed to become large [12]. This approximation makes the analysis tractable and is accurate for even small values of $M$ and $K$.

*Theorem 1*: For large $M$ and $K$, the sensitivity is:

$$\frac{\delta}{\varepsilon} = \frac{\mathrm{F}(\beta,\rho) - c_2(\beta,\rho)}{c_1(\beta,\rho)} \qquad (17)$$

where $c_1$ and $c_2$ are

$$c_1(\beta,\rho) = a(\log e)\left[\frac{\left(\sqrt{\beta}+1\right)^2}{4\sqrt{\beta}} + d\right] \qquad (18)$$

$$\begin{aligned}c_2(\beta,\rho) &= \beta\log\left(\frac{(1+\gamma)(1+\sqrt{1-a})}{2(\gamma+\sqrt{1-a})}\right) - (\log e)\frac{\left(\sqrt{\beta}-1\right)\left(1-\sqrt{1-a}\right)}{2(\gamma+\sqrt{1-a})} \\ &\quad - \frac{a(\log e)}{4}\left[-\left(\sqrt{\beta}+1\right) + 2d(\rho\beta - \rho - 1) + \frac{2\sqrt{\beta}}{\left(1+\sqrt{1-a}\right)^2}\right]\end{aligned} \qquad (19)$$

and

$$\begin{aligned}d &= \frac{\beta - 1}{2\sqrt{1-a}\left(1+\rho\left(\sqrt{\beta}+1\right)^2\right)(\gamma+\sqrt{1-a})} - \frac{\left(\sqrt{\beta}+1\right)^2}{2\sqrt{1-a}\left(1+\rho\left(\sqrt{\beta}+1\right)^2\right)\left(1+\sqrt{1-a}\right)^2} \\ &\quad - \frac{(\beta+1)}{2\left(1+\rho\left(\sqrt{\beta}+1\right)^2\right)\left(1+\sqrt{1-a}\right)^2}\end{aligned} \qquad (20)$$

and $a$ and $\gamma$ are defined in (16).

*Proof of Theorem 1*:

We take the derivatives of (14) with respect to $\beta$ and $\rho$:

$$\begin{aligned}I_{eq,K} &= \text{constant} \Rightarrow K\mathrm{F}(\beta,\rho) = \text{constant} \\ &\Rightarrow \frac{\partial(K\mathrm{F}(\beta,\rho))}{\partial\beta}d\beta + \frac{\partial(K\mathrm{F}(\beta,\rho))}{\partial\rho}d\rho = 0 \\ &\Rightarrow \left(\beta\frac{\partial\mathrm{F}(\beta,\rho)}{\partial\beta} - \mathrm{F}(\beta,\rho)\right)\frac{d\beta}{\beta} + \left(\rho\frac{\partial\mathrm{F}(\beta,\rho)}{\partial\rho}\right)\frac{d\rho}{\rho} = 0\end{aligned} \qquad (21)$$

Using (13) and (21), we obtain

$$\frac{\delta}{\varepsilon} = \frac{d\rho/\rho}{d\beta/\beta} = \left(\mathrm{F}(\beta,\rho) - \beta\frac{\partial\mathrm{F}(\beta,\rho)}{\partial\beta}\right) \bigg/ \left(\rho\frac{\partial\mathrm{F}(\beta,\rho)}{\partial\rho}\right) \qquad (22)$$

Therefore $c_2(\beta,\rho) = \beta \frac{\partial F(\beta,\rho)}{\partial \beta}$ and $c_1(\beta,\rho) = \rho \frac{\partial F(\beta,\rho)}{\partial \rho}$ and we omit the tedious derivative calculations. ∎

We notice that sensitivity $\delta/\varepsilon$ in (17) is a function of only $\rho$ and $\beta$ and is therefore "universal" in the sense that, on a complex Gaussian channel, it does not depend on the specific values of the number of transmit antennas $M$ and the number of users $K$ but only their ratio.

The sensitivity is the ratio of incremental power to user reduction while achieving constant rate. A low value of $\delta/\varepsilon$ implies that the rate is insensitive to the number of users being served; there is only a small power penalty if we serve fewer users. On the other hand, a large value of $\delta/\varepsilon$ implies that the rate is highly sensitive to the number of users being served.

Since the expression of sensitivity in (17) is rather complex, we look at some special cases and asymptotic results. For example, when $\beta = 1$ we obtain $\gamma = 0$ and we may simplify (17) to

$$\frac{\delta}{\varepsilon} = (\ln b)\left(1 + \frac{b}{\rho}\right) - \frac{1}{b}\left(1 + \frac{\rho}{b}\right) \qquad (23)$$

where $b = \left(1 + \sqrt{1+4\rho}\right)/2$.

We plot the sensitivity for $\beta = 1$ as given in (23) in Figure 1. One can see that the sensitivity is never negative because the mutual information is a non-decreasing function of the number of users being served. Furthermore, from Figure 1, the sensitivity for $\beta = 1$ can be separated into two regions, when $\rho$ is below 0 dB, the sensitivity $\delta/\varepsilon$ is small ($\delta/\varepsilon \ll 1$), but after $\rho = 0$ dB, $\delta/\varepsilon$ grows quickly. Consequently, at low SNR's the number of users can be decreased with only a small penalty ($\varepsilon$ large, $\delta$ small). However at high SNR a decrease in number of users can result in a large penalty in SNR ($\varepsilon$ small, $\delta$ large). In Figure 1, we also plot the sensitivity (17) for $\beta = 2, 4, 8$. We observe from the figure that the sensitivity increases as $\beta$ increases because a larger value of $\beta$ means that we are already serving fewer users than antennas.

We present the following asymptotics for the sensitivity. To a first-order approximation:

$\rho \to 0$:
$$\frac{\delta}{\varepsilon} = \frac{\beta\rho}{2} \qquad (24)$$

$\rho \to \infty$:
$$\frac{\delta}{\varepsilon} = \begin{cases} \ln(\rho)/2 - 1 & \text{if } \beta = 1 \\ \ln(\rho) + \ln(\beta - 1) - 1 & \text{if } \beta > 1 \end{cases} \qquad (25)$$

From (24) one can see that, at low $\rho$, the sensitivity is linear in $\rho$ with slope $\beta/2$ and goes to zero as $\rho$ goes to zero. This is seen in Figure 1. The effect of $\beta$ is to increase the penalty multiplicatively as $\beta$ increases.

When $\rho$ is large, (25) shows that there are two cases: $\beta = 1$ and $\beta > 1$. Hence the sensitivity as a function of $\ln(\rho)$ is significantly lower when $\beta = 1$ than otherwise. This suggests that we have much more freedom to reduce the number of users when we are serving the full $K = M$ than otherwise. The effect of $\beta > 1$ is to shift the curve to the left by $\ln(\beta - 1)$. This effect is also seen in Figure 1 in the curves for $\beta = 2, 4,$ and 8 at high SNR.

### 3.4 Application of sensitivity

Lets assume that $K = M$ ($\beta = 1$) and we ask: At what $\rho$ we can reduce the number of users to be served by ½, ¼, or ⅛ (equivalently, make $\beta' = 2, 4,$ or 8), while only suffering only $\approx 1$ dB power loss (keeping the achievable rate approximately constant)?

The sensitivity $\delta/\varepsilon$ can be used to answer this question. The complexity reduction coefficients $\varepsilon$ in (9) that correspond to changing $\beta$ from 1 to 2, or from 1 to 4, or from 1 to 8 are $\varepsilon = 1, 3,$ and 7 respectively. We choose the power penalty coefficient $\delta = 0.3$ since we then accept an estimated penalty of $10\log_{10}(1+0.3) \approx 1.1$ dB. When $\beta$ is changed from 1 to 2, we require a sensitivity of $\delta/\varepsilon = 0.3/1 = 0.3$, however when $\beta$ is changed from 1 to 8, we require a lower sensitivity of $0.3/8 = 0.043$.

We can obtain the operating point $\rho$ that yields these sensitivities by solving (23). Figure 2 is a plot of $\delta/\varepsilon$ versus $\rho$ that shows our operating points. For example, the sensitivity is 0.3 when $\rho = 1$ dB; this implies that when $\beta = 1$, there will be a loss of 1.1 dB at $\rho = 1$ dB when $\beta$ is increased to 2. A similar loss is obtained at $\rho = -6$ dB when $\beta$ is increased to 4; at $\rho = -10$ dB when $\beta$ is increased to 8. We may use these results to conclude that with $M = 8$ transmit antennas we can serve $K = 4$ users ($\beta = 2$) at 1 dB, $K = 2$ users ($\beta = 4$) at $-6$ dB, $K = 1$ user ($\beta = 8$) at $-10$ dB and still come within approximately 1.1 dB of $I_8$.

We verify these predicted operating points with the mutual information curves that are displayed in Figure 3. For example, Figure 3 shows that when $\rho = -6$ dB, in order to achieve the throughput of eight users ($\beta = 1$) but using only two users ($\beta$ becomes 4 and $\varepsilon = 3$), approximately 1.30 dB of extra power is needed. Similarly for the remaining two cases, it is found that they also require approximately 1.3 dB of extra power. Hence the (graphically) calculated penalties in $\rho$ are nearly the predicted 1.1 dB.

## 4. ALGORITHM PERFORMANCE WITH FEWER USERS

The previous sections show that we can reduce the number of users at low SNR and keep the sum-rate constant with only a small power penalty. We now show that reducing the number of users in some algorithms improves their performance and lowers their complexity sufficiently to overcome this penalty. For example, we examine the coded performance of $M = 8$ transmit antennas with $L = 8$ users at a total throughput of 8 bps/Hz using the vector-perturbation technique described in [9].

The vector-perturbation technique from [9] is summarized as follows:

$$\mathbf{y} = \frac{1}{\sqrt{\gamma}} \mathbf{H}\mathbf{H}^* \left( \mathbf{H}\mathbf{H}^* + \frac{\alpha}{\rho}\mathbf{I} \right)^{-1} (\mathbf{u} + \tau \mathbf{l}) + \mathbf{n} \qquad (26)$$

where $\tau = 2.5(|c|_{max} + \Delta/2)$, $\mathbf{y}$ is received signal vectors for all $K$ users, $\mathbf{H}$ is channel matrix, $\mathbf{u}$ and $\mathbf{n}$ are the $K$-dimensional signal and noise vectors for $K$ users, $\mathbf{G}$ is the regularized-inverse precoding matrix, $\gamma$ is a scalar such that total transmission power is normalized to one, $\alpha$ is the regularized-inverse parameter, $|c|_{max}$ is the absolute value of the constellation symbol with largest magnitude and $\Delta$ is the spacing between constellation points. The integer *perturbation vector $\mathbf{l}$* is obtained from the optimization

$$\mathbf{l} = \arg\min_{\mathbf{l}'} \left\{ (\mathbf{u} + \tau \mathbf{l}')^* (\mathbf{G}^*\mathbf{G})(\mathbf{u} + \tau \mathbf{l}') \right\} \qquad (27)$$

It is assumed that $\mathbf{u}$ consists of symbols coded with a rate-half turbo code from the UMTS standard with feedforward polynomial $1+D+D^3$, feedback polynomial $1+D^2+D^3$, block length 10000 bits, 20 inner iterations, and 8 outer loop iterations.

The BER performance of $M = 8$, $L = 8$, $K = 4$ and 8 users with a total throughput of 8 bps/Hz is shown in Figure 4. We observe that by serving $K = 4$ random users (dashed line, each user with 16QAM, hence rate 2 bps/Hz per user) can be better than serving $K = 8$ users (dotted line, each user with QPSK, hence rate 1 bps/Hz per user) even though the total data rate is the same. In this particular example, the gain is about 1.5 dB despite the fact that serving 4 users has a smaller channel sum-rate ($I_4$ = 8 bps/Hz at $\rho$ = 2.65 dB versus $I_8$ = 8 bps/Hz at $\rho$ = 1 dB). This illustrates the principle that serving fewer users at low SNR may lead to an algorithmic performance improvement that outweighs the power penalty. Furthermore, to serve four users we require only the CSI of any four users at a time, instead of all eight users, and thus the algorithm complexity decreases. We are approximately 3.2 dB away from $I_8$ by choosing four users randomly.

## 5. CONCLUSION

We provided a sensitivity study for a multiple-antenna multiple-user system of the number of users versus power while keeping the mutual information constant. Our study gave an analytical means for quantifying the power penalty for reducing the number of users served, especially at low SNR. We also showed that this loss can be compensated by improved algorithm performance and lower CSI requirements, and the user selection process can be fair.

Our results were universal in the sense that on a Gaussian channel they depended on only the ratio of the number of antennas to the number of users being served and could be applied to any number of antennas and users. One possible extension of this work would address users that have unequal SNR's and data-rate requirements. In this realistic scenario, the complexity of choosing how many users to serve (and at what rate) would probably be more difficult. Perhaps some combination of fairness and capacity-sensitivity would be needed to establish a good procedure.

### APPENDIX A - Proof of (7)

At low $\rho$, from (2) and (6) we know that:

$$I_M \geq I_{eq,M}$$
$$\mathrm{E} \max_{\mathbf{D},\mathrm{tr}(\mathbf{D})=1} \log \det\left(\mathbf{I}_M + \rho \mathbf{H}^*\mathbf{D}\mathbf{H}\right) \geq \mathrm{E} \log \det\left(\mathbf{I}_M + \frac{\rho}{M}\mathbf{H}^*\mathbf{H}\right) \tag{28}$$

where $\mathbf{H}$ is an $M \times M$ matrix.

When $\rho$ is small, the right hand side of (28) becomes:

$$I_{eq,M} = \mathrm{E} \log \det\left(\mathbf{I}_M + \frac{\rho}{M}\mathbf{H}^*\mathbf{H}\right) \approx (\log e)\frac{\rho}{M}\mathrm{Etr}\left(\mathbf{H}^*\mathbf{H}\right) = (\log e)\rho M \tag{29}$$

Similarly when $\rho$ is small, the left hand side of (28) becomes:

$$C_M = \mathrm{E} \max_{\mathbf{D},\mathrm{tr}(\mathbf{D})=1} \log \det\left(\mathbf{I}_M + \rho \mathbf{H}^*\mathbf{D}\mathbf{H}\right) \approx (\log e)\rho \mathrm{E} \max_{M\text{ i.i.d.}}\left(\frac{1}{2}\chi^2_{2M}\right) = (\log e)\rho \mathcal{M} \tag{30}$$

where $\mathcal{M} = \mathrm{E}\max\limits_{M\ \mathrm{i.i.d.}} \left(\frac{1}{2}\chi^2_{2M}\right)$ is the expected value of the maximum of $M$ i.i.d. chi-square random variables (normalized by ½), each with $2M$ degrees of freedom.

In fact, (30) suggests that at low $\rho$, the best strategy is to serve only the user with the maximum channel gain $\mathcal{M}$. We show that when $M$ is large, $\mathcal{M}$ is bounded as follows:

$$\lim_{M\to\infty} \Pr\left\{\frac{\mathcal{M}}{M} \leq (1+\zeta)\right\} = 1 \tag{31}$$

for any $\zeta > 0$. Then, by using (29), (30) and (31), we can bound (28) as

$$(\log e)\rho M(1+\zeta) \approx \mathrm{E}\max_{\mathbf{D},\mathrm{tr}(\mathbf{D})=1} \log\det(\mathbf{I}_M + \rho\mathbf{H}^*\mathbf{D}\mathbf{H}) \geq$$
$$\mathrm{E}\log\det\left(\mathbf{I}_M + \frac{\rho}{M}\mathbf{H}^*\mathbf{H}\right) \approx (\log e)\rho M \tag{32}$$

and we prove that lower bound in (6) and (28) are tight at low SNR.

To verify the inequality in (31), we need to examine the distribution of the maximum value, $\mathcal{M}$ of $M$ random independent $\frac{1}{2}\chi^2_{2M}$ variables.

$$\mathrm{P}(\mathcal{M} \leq x) = \left[\frac{1}{\Gamma(M)}\int_0^x e^{-t}t^{M-1}dt\right]^M \tag{33}$$

To find how $\mathcal{M}$ grows with $M$, it is suffices to find the smallest $x$ for which $\mathrm{P}(\mathcal{M} \leq x) \to 1$ as $M \to \infty$. We write:

$$\mathrm{P}(\mathcal{M} \leq x) = \left[1 - \frac{1}{\Gamma(M)}\int_x^\infty e^{-t}t^{M-1}dt\right]^M \tag{34}$$

For this probability to become 1 as $M \to \infty$, we need the integral to go to zero faster than $\mathrm{O}\left(\frac{1}{M}\right)$. For example, if the integral $\frac{1}{\Gamma(M)}\int_x^\infty e^{-t}t^{M-1}dt$ is $o\left(\frac{1}{M}\right)$ then $\ln \mathrm{P}(\mathcal{M} \leq x) \to 0$ or $\mathrm{P}(\mathcal{M} \leq x) \to 1$ as $M \to \infty$. We use

$$\frac{1}{\Gamma(M)}\int_x^\infty e^{-t}t^{M-1}dt = \frac{1}{\Gamma(M)}\int_0^\infty e^{-x(1+t')}x\left[x(1+t')\right]^{M-1}dt' \quad \text{where } t = x(1+t')$$
$$= \frac{1}{\Gamma(M)}e^{-x}x^M\int_0^\infty e^{-xt'}\left[(1+t')\right]^{M-1}dt' \tag{35}$$

Next, we use the approximation $1+t \leq e^t$ to yield

$$\frac{1}{\Gamma(M)}\int_x^\infty e^{-t}t^{M-1}dt \leq \frac{1}{\Gamma(M)}\frac{e^{-x}x^M}{x-(M-1)} \tag{36}$$

under the assumption that $x > M$-1. So we need to solve for $x > M$-1 such that

$$\frac{1}{\Gamma(M)}\frac{e^{-x}x^M}{x-(M-1)} = o\left(\frac{1}{M}\right) \tag{37}$$

as $M \to \infty$. Using the Stirling approximation for large $M$ [12] yields

$$\Gamma(M) \approx (M-1)^{M-1} e^{-(M-1)}\sqrt{2\pi(M-1)} \tag{38}$$

and taking the logarithm of (37) on both side, we obtain the requirement

$$M\ln M - \frac{1}{2}\ln M - M + \frac{1}{2}\ln(2\pi) + x - M\ln x + \ln(x-M+1) = \Omega(\ln M) \tag{39}$$

Let $x = M(1+\zeta)$, which satisfies $x > M$-1 for $\zeta > 0$. Then the left hand side of (39) becomes

$$\begin{aligned}&\frac{1}{2}\ln M + \frac{1}{2}\ln(2\pi) + M\zeta - M\ln(1+\zeta) + \ln\zeta \\ &= M\underbrace{\left(\zeta - \ln(1+\zeta)\right)}_{>0} + \frac{1}{2}\ln M + \frac{1}{2}\ln(2\pi) + \ln\zeta\end{aligned} \tag{40}$$

This satisfies the requirement because it is $\Omega(\ln M)$ for any $\zeta > 0$.

**List of Figures**

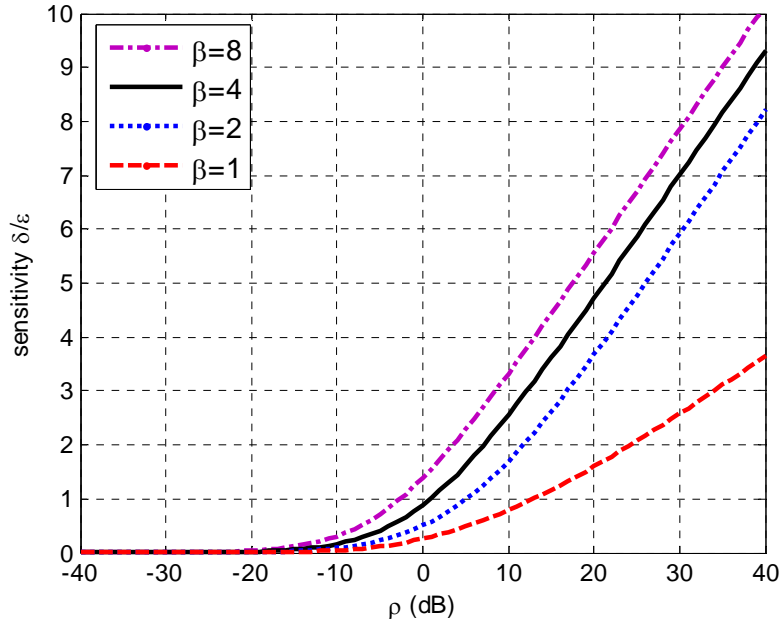

Figure 1: Sensitivity $\delta/\varepsilon$ for $\beta = 1, 2, 4, 8$. The curve $\beta = 1$, which is (23), can be separated into two parts: from $\rho =$ –40 dB to 0 dB, the sensitivity is small, but after $\rho = 0$ dB, $\delta/\varepsilon$ grows quickly. (Recall that small sensitivity implies that users may dropped while maintaining the sum-rate with little penalty in $\rho$.) The curves for $\beta = 2, 4, 8$ are given in (17). The curves grow rapidly as $\beta$ increases because we are already serving fewer users than antennas. The large-$\rho$ slope of these curves is given in (25).

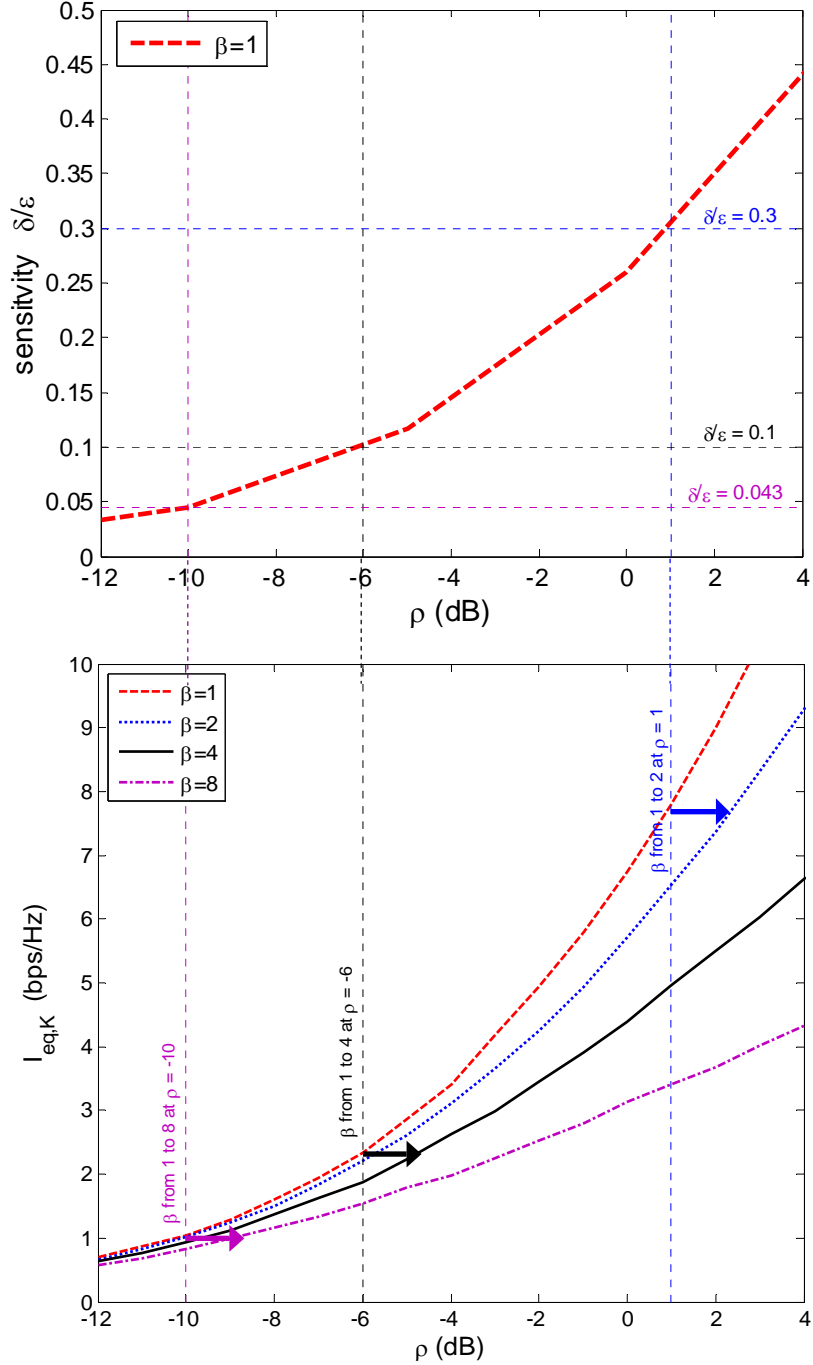

Figure 2: (upper figure) Sensitivity $\delta/\varepsilon$ for $\beta = 1$ as a function of $\rho$ (equation (23)). The markings at $\rho = 1$ dB, -6 dB, and -10 dB correspond to sensitivities of 0.3, 0.1 and 0.043. We are interested in $\delta = 0.3$, corresponding to an acceptable power penalty of 1.1 dB.

Figure 3: (lower figure) $I_{eq,K}$ for $M = 8$ transmit antennas with $\beta = 8, 4, 2, 1$ ($K = 1, 2, 4, 8$ users). As shown by the arrow at $\rho = 1$ dB, the power penalty for halving the number of users (increasing $\beta$ from 1 to 2 while keeping the mutual information constant) is approximately 1.4 dB, slightly larger than the design-point of 1.1 dB. The remaining two arrows show similar power penalties for increasing $\beta$ to 4 and 8.

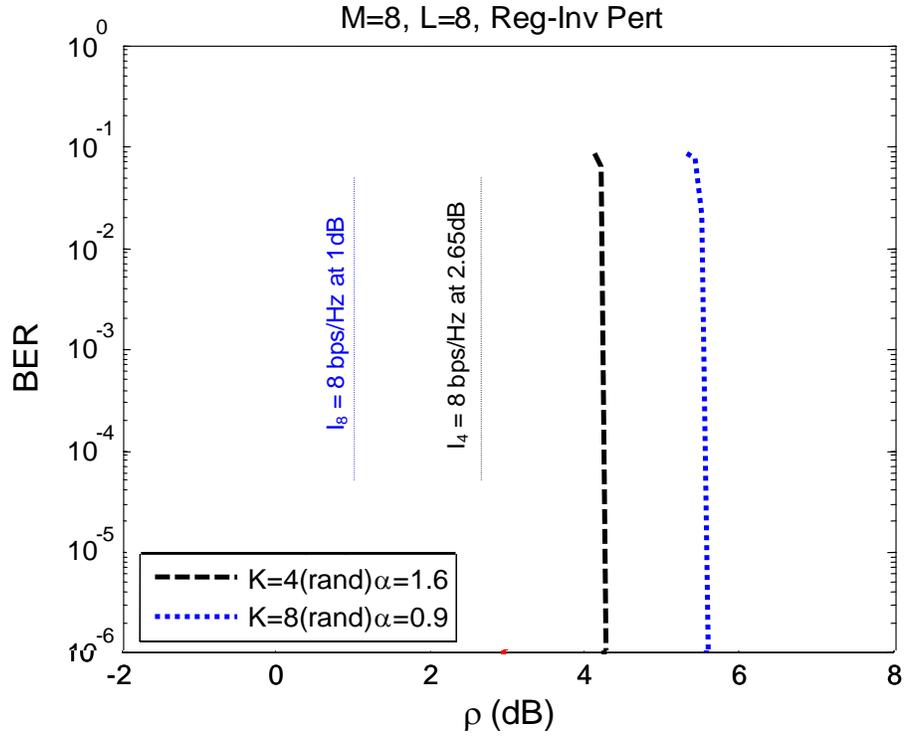

Figure 4: BER performance for $M = 8$ transmit antennas when serving $K = 4$ (dashed line) and 8 (dotted line) users at a throughput of 8 bps/Hz. The total pool of users is $L = 8$.